\documentclass[a4paper,11pt]{article}
\usepackage{cite}
\usepackage{fullpage,setspace,hyperref,comment,caption}
\usepackage[font=footnotesize,labelfont=bf,margin=1cm]{caption}
\usepackage{cite} 
\usepackage{color}
\usepackage{amsfonts}
\usepackage{amsmath}
\usepackage{amssymb}
\usepackage{todonotes} 
\usepackage{tikz}
\usepackage{tikz-cd}
\usepackage[utf8]{inputenc}
\usepackage{slashed}
\usepackage{tcolorbox}
\usepackage{bbm}
 \usepackage[normalem]{ulem}

\newcommand{\beq}{\begin{equation}}
\newcommand{\eeq}{\end{equation}}
\newcommand{\be}{\begin{equation}}
\newcommand{\ee}{\end{equation}}

\newcommand{\EFT}{ExFT}

\newcommand{\cA}{{\cal A}}
\newcommand{\cB}{{\cal B}}
\newcommand{\cC}{{\cal C}}
\newcommand{\CD}{{\cal D}}
\newcommand{\cE}{{\cal E}}

\newcommand{\obf}[1]{\overline{\mathbf{#1}}}
\newcommand{\mbf}[1]{\mathbf{#1}}

\numberwithin{equation}{section}

\newcommand{\LL}{\mathcal{L}}	

\newcommand{\gL}{\LL}

\newcommand{\SL}[1]{\mathrm{SL}( #1 )}

\newcommand{\Spin}[1]{\mathrm{Spin}(#1)}

\newcommand{\USp}[1]{\mathrm{USp}(#1)}
\newcommand{\EG}[1]{\mathrm{E}_{#1(#1)}}

\newcommand{\Odd}{\mathrm{O}(d,d)}

\newcommand{\Rone}{R_1}
\newcommand{\Rtwo}{R_2}

\newcommand\Tstrut{\rule{0pt}{3ex}}         
         
\newcommand\Bstrut{\rule[-1.3ex]{0pt}{0pt}}   
\usepackage{array}
\newcolumntype{H}{>{\setbox0=\hbox\bgroup}c<{\egroup}@{}}
 
\usepackage{tikz}
\usetikzlibrary{matrix,arrows,positioning,shapes,snakes,fadings,decorations.pathmorphing,
decorations.pathreplacing,automata,shadows,decorations.markings,patterns}
\usepackage[thicklines]{cancel}
\usetikzlibrary{calc,shapes.callouts,shapes.arrows}

            \usepackage{enumitem}

\allowdisplaybreaks[2]

\begin{document}
\begin{titlepage}

\vfill

\begin{center}
	\baselineskip=16pt  
	
	{\Large \bf  \it    $\EG{6}$ Exceptional Drinfel'd Algebras  }
	\vskip 2cm
	{\large \bf Emanuel Malek$^a$\footnote{\tt emanuel.malek@aei.mpg.de}, Yuho Sakatani$^b$\footnote{\tt yuho@koto.kpu-m.ac.jp} and Daniel C. Thompson$^{c,d}$\footnote{\tt d.c.thompson@swansea.ac.uk}}
	\vskip .6cm
	{\it $^a$ Max-Planck-Institut f\"{u}r Gravitationsphysik (Albert-Einstein-Institut), \\
		Am M\"{u}hlenberg 1, 14476 Potsdam, Germany \\ \ \\
		$^{b}$ Department of Physics, Kyoto Prefectural University of Medicine,\\
		Kyoto 606-0823, Japan \\  \ \\
					$^c$ Theoretische Natuurkunde, Vrije Universiteit Brussel, and the International Solvay Institutes, \\ Pleinlaan 2, B-1050 Brussels, Belgium \\ \ \\
			$^d$ Department of Physics, Swansea University, \\ Swansea SA2 8PP, United Kingdom \\ \ \\}
	\vskip 2cm

	\end{center}

\begin{abstract}
The exceptional Drinfel'd algebra (EDA) is a Leibniz algebra introduced to provide an algebraic underpinning with which to explore generalised notions of U-duality in M-theory. In essence, it provides an M-theoretic analogue of the way a Drinfel'd double encodes generalised T-dualities of strings. In this note we detail the construction of the EDA in the case where the regular U-duality group is $E_{6(6)}$.  We show how the EDA can be realised geometrically as a generalised Leibniz parallelisation of the exceptional generalised tangent bundle for a six-dimensional group manifold $G$, endowed with a Nambu-Lie structure.  When the EDA is of coboundary type, we show how a natural generalisation of the classical Yang-Baxter equation arises. 
The construction is illustrated with a selection of examples including some which embed Drinfel'd doubles and others that are not of this type. 
\end{abstract}

\vfill

\setcounter{footnote}{0}
\end{titlepage}
 \tableofcontents 
 \newpage
 
  \section{Introduction}
 Dualities play an important role in our understanding of string theory. One of the best-understood dualities is T-duality, which relates string theory on backgrounds with $U(1)^{d}$ isometries, with the backgrounds related by $\Odd$ transformations. These T-dualities are already visible in perturbative string theory, and are enlarged into U-dualities in the non-perturbative framework of M-theory \cite{Hull:1994ys,Witten:1995ex}. Generalisations of Abelian T-dualities exist for backgrounds with non-Abelian isometries, leading to non-Abelian T-duality (NATD) \cite{delaOssa:1992vci}, and for backgrounds without any isometries, called Poisson-Lie T-duality (PLTD) \cite{Klimcik:1995dy,Klimcik:1995ux}. Instead of the isometry algebra, PLTD is controlled by an underlying Drinfel'd double.

Unlike Abelian T-duality, which is an equivalence between string theories on different backgrounds to all orders in the string coupling and string length, these generalised T-duality are currently best understood at the supergravity level and only to a limited extend beyond leading order in $\alpha'$ \cite{Hoare:2019mcc} and their status as true dualities of the   string genus expansion remains doubtful \cite{Giveon:1993ai}. Nonetheless, NATD and PLTD have led to fruitful results. For example, NATD has been successfully used as solution-generating mechanisms of supergravity \cite{ Sfetsos:2010uq}, leading to the discovery of new minimally supersymmetric AdS backgrounds starting with \cite{Itsios:2013wd } (see \cite{Thompson:2019ipl} for a review and further references). Moreover, there is a close connection between PLTD and the (modified) classical Yang-Baxter equation which controls integrable deformations of $\sigma$-models \cite{Klimcik:2002zj,Klimcik:2008eq}.

The non-perturbative generalisation of Poisson-Lie T-duality to a U-duality version in M-theory, or more conservatively as a solution-generating mechanism of 11-dimensional supergravity, has long been an open problem, which was recently addressed in \cite{Sakatani:2019zrs,Malek:2019xrf} and further elaborated on  in \cite{Sakatani:2020iad,Hlavaty:2020pfj,Blair:2020ndg}. Building on the interpretation of PLTD and Drinfel'd doubles within Double Field Theory (DFT), \cite{Hassler:2017yza,Demulder:2018lmj,Sakatani:2019jgu}, \cite{Sakatani:2019zrs,Malek:2019xrf} used Exceptional Field Theory (ExFT)/Exceptional Generalised Geometry to propose a natural generalisation of the Drinfel'd double for dualities along four spacetime dimensions. This ``Exceptional Drinfel'd Algebra'' (EDA) was shown to lead to a new solution-generating mechanism of 11-dimensional supergravity that suggests a notion of Poisson-Lie U-duality, as well as a generalisation of the classical Yang-Baxter equation. Other recent works  \cite{Bakhmatov:2019dow,Bakhmatov:2020kul,Musaev:2020bwm} have considered closely related ideas, although the detailed relation between these approaches and the EDA is not completely apparent. 

In this paper, we will further develop the ideas of \cite{Sakatani:2019zrs,Malek:2019xrf} by constructing EDAs and Poisson-Lie U-duality amongst six directions. We choose six dimensions, because important new features arise when dualities are considered in six directions. This is because now the 6-form can completely wrap the six directions we are considering. As a result, the $\frak{e}_{6(6)}$ algebra contains a generator, corresponding to a hexavector, which will generate new kinds of dualities and deformations which have no counterpart in PLTD, as we will see. 

The outline of the rest of this paper is as follows: in section \ref{s:LinCon} we describe the EDA from a purely algebraic perspective.  In section \ref{s:GenFrame} we show how the EDA can be realised within exceptional generalised geometry as a Leibniz parallelisation of a particular type of group manifold $G$, that we will call a $(3,6)$-Nambu-Lie group.  We then consider more closely the case of a coboundary EDA  in section \ref{s:YB} whose structure is governed by a generalisation of the Yang Baxter equation. We provide a range of examples in section \ref{s:Examples} of EDAs both coboundary and otherwise, some of which have   Drinfel'd doubles as subalgebras, and other which do not have such an interpretation.  The aim of these examples is not to provide here a full classification, which could form an interesting investigation in its own right, but rather to highlight the various features that can arise.

\section{The $E_{6(6)}$ EDA} \label{s:LinCon} 
Before specialising to the case of $E_{6(6)}$ we begin by presenting some generalities of the Exceptional Drinfel'd Algebra.   
The EDA, $\frak{d}_n$, is a Leibniz algebra which is a subalgebra of $E_{n(n)}$ \footnote{In general one can allow for EDAs as subalgebras of $E_{n(n)} \times \mathbb{R}^+$, see \cite{Malek:2019xrf}. However, we will not deal with the extra $\mathbb{R}^+$ factor here.}, admitting a ``maximally isotropic'' subalgebra, as we will define shortly.  In table 1 we provide details of the representations of   $E_{n(n)}$    inherited from the exceptional field theory (ExFT) approach to eleven-dimensional supergravity that are useful to the present construction. 
  
 \renewcommand{\arraystretch}{1.1}
\begin{table}[h]\centering
			\begin{tabular}{|c|c|c|c|c|c|c H c|}
				\hline
				$D$ & $E_{n(n)}$ & $H_n$ & $R_1$ & $R_2$ & $R_3$ &$R_4$ & $R_c$ & \Tstrut\Bstrut \\ \hline 
				7 & $\SL{5}$ & $\USp{4} / \mathbb{Z}_2$ & $\mbf{10}$ & $\obf{5}$ & $\mbf{5}$ &$\obf{10}$  & $\emptyset$ & \\
				6 & $\Spin{5,5}$ & $\USp{4}\times\USp{4} / \mathbb{Z}_2$ & $\mbf{16}$ & $\mbf{10}$ & $\obf{16}$ & $\mbf{45}$ & $\mbf{1}$ &\\
				5 & $\EG{6}$ & $\USp{8} / \mathbb{Z}_2$  & $\mbf{27}$ & $\obf{27}$ & $\mbf{78}$ &$\obf{351'}$ & $\mbf{27}$ &\\
				\hline
			\end{tabular}
			\vskip-0.5em
			\captionof{table}{\small{The split real form of   exceptional groups $E_{n(n)}$  with $D=11-n$, their maximal compact subgroups $H_n$ and representations $R_1 \dots R_4$ appearing in the tensor hierarchy of ExFT. 
			In this work we will be mostly concerned with representations $R_1$ and $R_2$ which will be associated to the generalised tangent bundles $E$ and $N$ respectively.}}  \label{t:Edd}
\end{table}
\renewcommand{\arraystretch}{1}

We  denote the generators of $\frak{d}_n$ by $\{ T_A \}$, with the index $A$ inherited from the $\bar{R}_1$ representation of $E_{n(n)}$ {\EFT} and their product by
 \begin{align}
T_A \circ T_B  =  X_{AB}{}^C\, T_C\,,
\end{align}
with $X_{AB}{}^C$ structure constants which are not necessarily antisymmetric in their lower indices. The product obeys the Leibniz identity, namely 
\begin{align} \label{eq:Leibniz}
T_A \circ (T_B \circ T_C) = (T_A\circ T_B)\circ T_C + T_B\circ( T_A \circ T_C)\, , 
\end{align}
which implies for the structure constants
\begin{align}\label{eq:QC1}
 X_{AC}{}^D{}\,X_{BD}{}^E - X_{BC}{}^D\,X_{AD}{}^E + X_{AB}{}^D\,X_{DC}{}^E = 0 \,.
\end{align} 
Note that if the Leibniz algebra is a Lie algebra, i.e. the $X_{AB}{}^C$ are antisymmetric in their lower indices, then this reduces to the Jacobi identity.

We place two further (linear) requirements on the EDA. Firstly, we demand that there is a maximal \underline{Lie}   subalgebra $\frak{g}$ spanned by $\{T_a\}\subset \{ T_A \}$ obeying 
\begin{align}\label{eq:seccond}
\frak{g} \otimes \frak{g}|_{\bar{R}_2} = 0 \,, 
\end{align}
in which the representation $ \bar{R}_2$ is found in table 1. We call such a subalgebra $\frak{g}$ maximally isotropic. We will be interested here in the case that $\dim \frak{g} = n$ as this is relevant to the M-theory context.\footnote{There is another inequivalent way to maximally solve the condition eq. \eqref{eq:seccond} with $\dim \frak{g} = n-1$ leading to a IIB scenario \cite{Blair:2013gqa,Hohm:2013vpa}.}   Since $G =\exp \frak{g}$   acts adjointly on $\frak{d}_n$, it follows that $G$ should be  endowed with a trivector and hexavector.  We will further require that these objects give rise to a 3- and 6-bracket on $\frak{g}^*$, thereby imposing some further restrictions on the structure constants $X_{AB}{}^C$.\footnote{It is worth emphasising that these are impositions beyond simply demanding that $\frak{g}$ be a maximal isotropic.}  These additional requirements imply that the EDA can be given a geometrical realisation in terms of certain generalised frames whose action is mediated by the  generalised Lie derivative \eqref{eq:gLie}, as we will show in section \ref{s:GenFrame}.

Let us now discuss these restrictions in detail.
 
 \subsection{Linear Constraints}
We now study in detail the consequence of the requirements of the maximally isotropic subalgebra $\frak{g}$ and its adjoint action.  Since these constraints arise from placing requirements directly to  the form of $X_{AB}{}^C$ we describe them as linear constraints; this is to be contrasted with quadratic constraints of the form $X^2=0 $ that arise from the Leibniz identity.    

Firstly, since $\frak{g}$ is a Lie algebra, we immediately have
\begin{equation}
    \begin{split}
         T_a \circ T_b &= f_{ab}{}^c\,T_c \,,
    \end{split}
\end{equation}
with $f_{ab}{}^c$ antisymmetric in $a$, $b$. Secondly, the adjoint action\footnote{To be more precise we inherit an   action via the rack product:
\begin{align*}
    g \cdot T_A \cdot g^{-1} \equiv g^{-1} \triangleright T_A \equiv T_A + h \circ T_A + \frac{1}{2}  h \circ (h \circ T_A) + \cdots \qquad (g^{-1}\equiv e^h)\,. 
\end{align*}}
of $g \in G = \exp{\frak{g}}$ on $\frak{d}_n$ implies that
\begin{equation}
    g \cdot T_A \cdot g^{-1} = \left(A_g\right)_A{}^B T_B \,,
\end{equation}
with $\left(A_g\right)_A{}^B \in E_{6(6)}$ since $\mathfrak{g} \subset \frak{d}_n \subset \frak{e}_{6(6)}$. Let us denote the adjoint action of $g \in G$ on $\frak{g}$ by $a_g$.

Then $g \cdot T_A \cdot g^{-1}$ takes the form:
\begin{equation}
    \begin{split} \label{eq:adjoint}
        g \cdot T_a \cdot g^{-1} &= (a_g)_a{}^b\,T_b\,,\\
        g \cdot T^{a_1a_2} \cdot g^{-1} &= - \lambda_g^{a_1a_2c}\,(a_g)_c{}^b\,T_b + (a_g^{-1})_{b_1}{}^{a_1}\,(a_g^{-1})_{b_2}{}^{a_2}\,T^{b_1b_2}\,,\\
        g \cdot T^{a_1\ldots a_5} \cdot g^{-1} &= \bigl(\lambda_g^{a_1\ldots a_5c}+ 5\,\lambda_g^{[a_1a_2a_3}\,\lambda_g^{a_4a_5]c}\bigr)\,(a_g)_c{}^b\,T_b \\
        &\quad - 10\,\lambda_g^{[a_1a_2a_3}\,(a_g^{-1})_{b_1}{}^{a_4}\,(a_g^{-1})_{b_2}{}^{a_5]}\,T^{b_1b_2} \\
        &\quad + (a_g^{-1})_{b_1}{}^{[a_1}\ldots (a_g^{-1})_{b_5}{}^{a_5]}\,T^{b_1\ldots b_5}\,.
    \end{split}
\end{equation}
Thus, $G$ admits a totally antisymmetric trivector $\lambda^{abc}$ and totally antisymmetric hexavector $\lambda^{a_1\ldots a_6}$ which control its adjoint action on the generators $T^{ab}$ and $T^{a_1 \ldots a_5}$.

Equations \eqref{eq:adjoint} imply that $\left(\lambda_g\right)^{abc}$ and $\left(\lambda_g\right)^{a_1 \ldots a_6}$ vanish at the identity, i.e.
\begin{equation} \label{eq:LambdaE}
    \left(\lambda_e\right)^{abc} = \left( \lambda_e\right)^{a_1 \ldots a_6} = 0 \,,
\end{equation}
and they inherit a group composition rule
\begin{equation}
    \begin{split} \label{eq:LambdaComp}
        \lambda_{hg}^{a_1a_2a_3} 
        &= \lambda_g^{a_1a_2a_3} 
        + (a_g^{-1})_{c_1}{}^{a_1}\,(a_g^{-1})_{c_2}{}^{a_2}\,(a_g^{-1})_{c_3}{}^{a_3}\, \lambda_h^{c_1c_2c_3} \,, \\
        \lambda_{hg}^{a_1\ldots a_6} 
        &= \lambda_{g}^{a_1\ldots a_6} + (a_{g}^{-1})_{c_1}{}^{a_1}\ldots (a_g^{-1})_{c_6}{}^{a_6}\, \lambda_{h}^{c_1\ldots c_6} \\
        &\quad + 10\,\lambda_g^{[a_1a_2a_3}\,(a_{g}^{-1})_{c_1}{}^{a_4}\,(a_{g}^{-1})_{c_2}{}^{a_5}\,(a_g^{-1})_{c_3}{}^{a_6]}\, \lambda_h^{c_1c_2c_3}\,,
    \end{split}
\end{equation}
for $g, h \in G$.

Finally, we come to the second condition on the EDA, i.e. the existence of a 3- and 6-bracket on $\frak{g}^*$. This is equivalent to imposing the following differential conditions on $\lambda^{abc}$ and $\lambda^{a_1\ldots a_6}$:  
\begin{equation}
    \begin{split} \label{eq:dlambda}
        d\lambda^{a_1a_2a_3} &=   r^b \left( f_b{}^{a_1a_2a_3} + 3\,f_{bc}{}^{[a_1} \, \lambda^{|c|a_2a_3]} \right) \,, \\
        d\lambda^{a_1\ldots a_6} &= r^b \left( f_b{}^{a_1\ldots a_6} + 6\,f_{bc}{}^{[a_1}\, \lambda^{|c|a_2\ldots a_6]} + 10\, f_b{}^{[a_1a_2a_3}\, \lambda^{a_4a_5a_6]} \right) \,,
    \end{split}
\end{equation}
where $r =    r^a\, T_a$ are the right-invariant 1-forms on $G$ obeying $dr^a = \frac{1}{2} f_{bc}{}^a r^b \wedge r^c$  and we have dropped the subscript $g$ on $\lambda^{(3)}$ and $\lambda^{(6)}$.  The $f_b{}^{a_1 \ldots a_3}$ and $f_{b}{}^{a_1 \ldots a_6}$ are structure constants for a 3- and 6-bracket and are totally antisymmetric in their upper indices.  In fact, as we will see in section \ref{s:QC}, the Leibniz identity implies further properties of the trivector and hexavector, in particular that they define a certain Nambu 3- and 6-bracket which are compatible with the Lie bracket on $G$. Therefore, it seems apt to call $G$ a (3,6)-Nambu-Lie Group.

With the above conditions, the EDA takes the following form 
\begin{equation}\label{eq:TheEDA}
    \begin{split}
     T_a \circ T_b &= f_{ab}{}^c\,T_c \,, \\
     T_a \circ T^{b_1b_2} &= f_a{}^{b_1b_2c}\,T_c + 2\,f_{ac}{}^{[b_1}\,T^{b_2]c}\,,\\
     T_a \circ T^{b_1\ldots b_5} &= -f_a{}^{b_1\ldots b_5c}\,T_c + 10\,f_{a}{}^{[b_1b_2b_3}\,T^{b_4b_5]} - 5\,f_{ac}{}^{[b_1}\,T^{b_2\ldots b_5]c} \,, \\
    T^{a_1a_2} \circ T_b &= -f_b{}^{a_1a_2c}\,T_c + 3\,f_{[c_1c_2}{}^{[a_1}\,\delta^{a_2]}_{b]}\,T^{c_1c_2}\,,\\
    T^{a_1a_2} \circ T^{b_1b_2} &= -2\, f_c{}^{a_1a_2[b_1}\, T^{b_2]c} + f_{c_1c_2}{}^{[a_1}\,T^{a_2]b_1b_2c_1c_2}\,,\\
    T^{a_1a_2} \circ T^{b_1\ldots b_5} &= 5\,f_c{}^{a_1a_2[b_1}\, T^{b_2\ldots b_5]c} \,, \\
     T^{a_1\ldots a_5} \circ T_b &= f_b{}^{a_1\ldots a_5c}\,T_c - 10\,f_b{}^{[a_1a_2a_3}\,T^{a_4a_5]} -20\,f_c{}^{[a_1a_2a_3}\,\delta_b^{a_4}\,T^{a_5]c} \\
     &\quad + 5\,f_{bc}{}^{[a_1}\,T^{a_2\ldots a_5]c} + 10\,f_{c_1c_2}{}^{[a_1}\,\delta^{a_2}_b\,T^{a_3a_4a_5]c_1c_2} \,, \\
    T^{a_1\ldots a_5} \circ T^{b_1b_2} &= 2\,f_c{}^{a_1\ldots a_5[b_1}\,T^{b_2]c} - 10\,f_c{}^{[a_1a_2a_3}\, T^{a_4a_5]b_1b_2c}\,, \\
    T^{a_1\ldots a_5} \circ T^{b_1\ldots b_5} &= -5\,f_c{}^{a_1\ldots a_5[b_1}\, T^{b_2\ldots b_5]c} \,.
  \end{split}
\end{equation}
\subsection{Leibniz identity constraints} \label{s:QC}
We will now study the compatibility conditions between the Lie algebra $\frak{g}$, the 3-bracket and 6-bracket, as well as their appropriate ``closure'' conditions that are required for the EDA to satisfy the Leibniz identity of eq. \eqref{eq:Leibniz}.
This yields a number of immediate  constraints. In particular, we obtain the following fundamental identities, i.e. generalisations of Jacobi for higher brackets,
\begin{align} 
    \label{eq:FI-1} 
         0  &= 3 f_{[a_1 a_2}{}^c \,f_{a_3]c}{}^b   \,, \\  
    \label{eq:FI-3}
       0 & =  f_a{}^{d c_1c_2}\, f_d{}^{b_1b_2b_3} - 3\, f_d{}^{c_1c_2[b_1}\,f_a{}^{b_2b_3]d} + f_{d_1d_2}{}^{[c_1}\, f_a{}^{c_2] b_1b_2b_3 d_1d_2}   \,,  \\ 
    \label{eq:FI-6}
        0 &=  f_a{}^{d c_1\ldots c_5}\, f_d{}^{b_1\ldots b_6} - 6\,f_d{}^{c_1\ldots c_5[b_1} f_a{}^{b_2\ldots b_6]d}  \,,
\end{align}
as well as compatibility conditions between the dual structure constants $f_b{}^{a_1a_2 a_3}$, $f_b{}^{a_1\ldots a_6}$ and the Lie algebra structure constants. These compatibility conditions take the form of cocycle conditions 
\begin{align}
    \label{eq:cocycle-3}
        0 &= f_{a_1a_2}{}^c\,f_c{}^{b_1b_2b_3} + 6\, f_{c[a_1}{}^{[b_1|}\, f_{a_2]}{}^{c|b_2b_3]}   \,, \\ 
    \label{eq:cocycle-6}
        0 &= f_{a_1a_2}{}^c\,f_c{}^{b_1\ldots b_6} +12\,f_{c[a_1}{}^{[b_1|}\,f_{a_2]}{}^{c|b_2\ldots b_6]} -20\, f_{[a_1}{}^{[b_1b_2b_3}\,f_{a_2]}{}^{b_4b_5b_6]}   \,,
\end{align} 
as well as the additional constraint
 \begin{equation} \label{eq:addcon}
 f_{d_1 d_2}{}^a f_{c}{}^{d_1 d_2 b} = 0 \,.  
\end{equation}

If we only consider EDAs   $\frak{d}_n$ with $n\leq6$, as we are doing here, the   conditions given by the above eqs. \eqref{eq:FI-1}-\eqref{eq:addcon} are equivalent to imposing the Leibniz identity. This is because in $n \leq 6$, the fundamental identity for the six-bracket implies that $f_{b}{}^{a_1 \ldots a_6} = 0$. However, since the structure we are studying here will also exist for $n > 6$, we will keep the remaining discussion as dimension-independent as possible, whilst keeping in mind that for $n > 6$, the Leibniz identity will lead to further or modified compatibility conditions between $f_{ab}{}^c$, $f_b{}^{a_1 a_2 a_3}$ and $f_{b}{}^{a_1 \ldots a_6}$. These additional constraints will need to be studied using EDAs based on $E_{7(7)}$ and higher.

Before interpreting these constraints, we remark that the Leibniz identity ensures, much as the structure constants of a Lie algebra $\frak{g}$ are  invariant under    $G=\exp\frak{g}$ acting adjointly, that the  EDA structure constants enjoy an invariance 
\begin{equation}\label{eq:adinvariance}
  X_{AB}{}^{D} (A_g)_D{}^C = (A_g)_A{}^D (A_g)_B{}^E X_{DE}{}^C \, .  
\end{equation}
Substitution of eq. \eqref{eq:adjoint} here results in a variety of identities that we shall revisit later on.

\subsubsection{Fundamental identities}
Let us now introduce the 3-bracket $\left\{\, \right\}_3$ and 6-bracket $\left\{\, \right\}_6$ on $\frak{g}^*$ with structure constants $f_b{}^{a_1 a_2 a_3}$ and $f_b{}^{a_1 \ldots a_6}$, respectively, i.e.
\begin{equation}
    \begin{split}
        \left\{ x,\, y,\, z \right\}_3 &= f_a{}^{b_1b_2b_3}\, x_{b_1}\, y_{b_2}\, z_{b_3} \,, \\
        \left\{ u,\, v,\, w,\, x,\, y,\, z \right\}_6 &= f_a{}^{b_1\ldots b_6}\, u_{b_1}\, v_{b_2}\, w_{b_3}\, x_{b_4}\, y_{b_5}\, z_{b_6} \,.
    \end{split}
\end{equation}
The conditions \eqref{eq:FI-1} and \eqref{eq:FI-3} imply that the 3- and 6-brackets satisfy
\begin{equation}
    \begin{split} \label{eq:FIa}
     \left\{ x_1,\, x_2,\, \left\{ x_3,\, x_4,\, x_5 \right\}_3 \right\}_3 &= \left\{ \left\{x_1,\, x_2,\,x_3 \right\}_3,\, x_4,\, x_5\right\}_3 + \left\{ x_3,\, \left\{x_1,\, x_2,\,x_4 \right\}_3,\, x_5\right\}_3 \\
     & \quad + \left\{ x_3,\, x_4,\, \left\{x_1,\, x_2,\,x_5 \right\}_3 \right\}_3 \\
     &- \left\{ \Delta(x_1), x_2, x_3, x_4, x_5 \right\}_6 + \left\{ \Delta(x_2), x_1, x_3, x_4, x_5 \right\}_6 \,, \\
     \left\{ x_1,\, \ldots ,\, x_5,\, \left\{ y_1,\, \ldots ,\, y_6 \right\}_6 \right\}_6 &= \left\{ \left\{ x_1,\,\ldots ,\, x_5,\, y_1 \right\}_6,\, y_2,\, \ldots,\, y_6 \right\}_6 \\
     & \quad + \left\{ y_1,\, \left\{ x_1,\, \ldots,\, x_5,\, y_2 \right\}_6,\, y_3,\, \ldots,\, y_6 \right\}_6 \\
     & \quad + \left\{ y_1,\, y_2,\, \left\{ x_1,\, \ldots,\, x_5,\, y_3 \right\}_6,\, y_4,\, \ldots,\, y_6 \right\}_6 \\
     & \quad + \left\{ y_1,\, \ldots,\, y_3,\, \left\{ x_1,\, \ldots,\, x_5,\, y_4 \right\}_6,\, y_5,\, y_6 \right\}_6 \\
     & \quad + \left\{ y_1,\, \ldots,\, y_4,\, \left\{ x_1,\, \ldots,\, x_5,\, y_5 \right\}_6,\, y_6 \right\}_6 \\
     & \quad + \left\{ y_1,\, \ldots,\, y_5,\, \left\{ x_1,\, \ldots,\, x_5,\, y_6 \right\}_6 \right\}_6 \,,
    \end{split}
\end{equation}
for all $x_1,\, \ldots,\, x_5,\, y_1,\, \ldots,\, y_6  \in \frak{g}^*$, and where we used the Lie bracket on $\frak{g}$ to define the $\textrm{ad}$-invariant co-product $\Delta$ on $\frak{g}^*$
\begin{equation}
    \begin{split} \label{eq:adcoproduct}
    \Delta: \frak{g}^* &\longrightarrow \frak{g}^* \wedge \frak{g}^* \,, \\
    ad_x \Delta(y) &= \Delta(ad_x y) \,,\,\, \forall\; x \in \frak{g},\, y \in \frak{g}^* \,,
    \end{split}
\end{equation}
which is given, assuming a basis $\{T^a\}$ for $\frak{g}^*$, by
\begin{equation}
    \begin{split} \label{eq:adcoproductind}
        \Delta(x_a\, T^a) = \frac12 f_{bc}{}^{a}\, x_a\, T^b \wedge T^c \,.
    \end{split}
\end{equation} 
We see that the 6-bracket must satisfy the fundamental identity for Nambu 6-brackets, while the 3-bracket's fundamental identity is modified by the 6-bracket and the co-product defined by the structure constants of $\frak{g}$. 

\subsubsection{Compatibility conditions}
The first set of compatibility conditions, eqs. \eqref{eq:cocycle-3} and \eqref{eq:cocycle-6},  between the 3- and 6-brackets and the Lie algebra $\frak{g}$ imply that $f_b{}^{a_1a_2a_3}$ defines a $\frak{g}$-cocycle and that $f_b{}^{a_1\ldots a_6}$ is an $f_3$-twisted $\frak{g}$-cocycle, as follows. $f_b{}^{a_1a_2a_3}$ and $f_b{}^{a_1\ldots a_6}$ define $\Lambda^3 \frak{g}$- and $\Lambda^6 \frak{g}$-valued 1-cochains
\begin{equation}
    \begin{split}
        f_3: \frak{g} &\longrightarrow \Lambda^3 \frak{g} \,, \\
        f_6: \frak{g} &\longrightarrow \Lambda^6 \frak{g} \,,
    \end{split}
\end{equation}
defined by
\begin{equation}
    \begin{split}
        f_3(x) &= \frac1{3!} x^b\, f_b{}^{a_1a_2a_3}\, T_{a_1} \wedge T_{a_2} \wedge T_{a_3} \,,\; \forall\, x = x^a T_{a} \in \frak{g} \,, \\
        f_6(x) &= \frac{1}{6!} x^b\, f_b{}^{a_1 \ldots a_6}\, T_{a_1} \wedge \ldots \wedge T_{a_6} \,,\; \forall\, x = x^a T_{a} \in \frak{g} \,.
    \end{split}
\end{equation}
Using the coboundary operator $d:\frak{g}^*\otimes\Lambda^p\frak{g} \longrightarrow \Lambda^2\frak{g}^*\otimes\Lambda^p\frak{g}$, for $p=3$ and $p=6$ here,
\begin{align}
 df_3(x,y)&\equiv ad_x f_3(y) -ad_y f_3(x) - f_3([x,y])\,,
\\
 df_6(x,y)&\equiv ad_x f_6(y) -ad_y f_6(x) - f_6([x,y])\, ,
\end{align}
the conditions \eqref{eq:cocycle-3} and \eqref{eq:cocycle-6} are more elegantly stated as  
\begin{align}\label{eq:cocycle-condensed}
 df_3(x,y) = 0\,,\qquad 
 df_6(x,y) + f_3(x)\wedge f_3(y) = 0\,. 
\end{align}

The coboundary operator is nilpotent with $d:\Lambda^p\frak{g}\longrightarrow\frak{g}^*\otimes\Lambda^p\frak{g}$ defined as
\begin{equation}
    d\rho_p(x) = ad_x \rho_p \,,
\end{equation}
for all $x \in \frak{g}$ and $\rho_p \in \Lambda^p\frak{g}$. Therefore, the cocycle conditions \eqref{eq:cocycle-condensed} can be solved by the (twisted) coboundaries
\begin{equation}
    \begin{split} 
        f_3 &= d\rho_3 \,, \qquad f_6 = d\rho_6 + \frac12 \rho_3 \wedge d\rho_3 \,.
    \end{split}
\end{equation}
In components, these are equivalent to
\begin{equation}
    \begin{split}
        f_a{}^{b_1b_2b_3} &= 3\,f_{ac}{}^{[b_1}\,\rho^{|c|b_2b_3]}\,, \\
        f_a{}^{b_1\ldots b_6} &= 6\,f_{ac}{}^{[b_1|}\,\rho^{c|b_2\ldots b_6]} + 30\,f_{ac}{}^{[b_1}\,\rho^{|c|b_2b_3}\,\rho^{b_4b_5b_6]} \,. \label{eq:coboundary}
    \end{split}
\end{equation}
The coboundary case is related to a generalisation of Yang-Baxter deformations. The trivector $\rho^{a_1a_2a_3}$ and the hexavector $\rho^{a_1\ldots a_6}$ correspond to the M-theoretic analogue of the classical $r$-matrix. The equations corresponding to the classical Yang-Baxter equations for the $r$-matrices are implied by substituting the solutions \eqref{eq:coboundary} to the fundamental identities \eqref{eq:FI-3} and  \eqref{eq:FI-6}. We will discuss this further in section \ref{s:YB}.

Finally, the additional constraint \eqref{eq:addcon} implies that the ad-invariant co-product $\Delta$ on $\frak{g}^*$ \eqref{eq:adcoproduct} defines a commuting subspace of the 3-bracket:
\begin{equation}
    \left\{ \Delta(x_1),\, x_2 \right\}_3 = 0 \,,\; \forall\, x_1, x_2 \in \frak{g}^*\,.
\end{equation}

\section{$E_{6(6)}$ EDA from generalised frame fields} 
\label{s:GenFrame} 
We now provide a geometric realisation of the $E_{6(6)}$ EDA by constructing a Leibniz parallelisation \cite{Grana:2008yw,Aldazabal:2011nj,Geissbuhler:2011mx,Berman:2012uy,Lee:2014mla,Hohm:2014qga} of the exceptional generalised tangent bundle \cite{Pacheco:2008ps,Berman:2010is,Berman:2011cg,Coimbra:2011ky,Berman:2012vc,Coimbra:2012af,Hohm:2013pua}  
  \begin{align}
 E &\cong TM\oplus \Lambda^2 T^*M \oplus \Lambda^5 T^*M \, , 
 \end{align}
 in which we identify the manifold $M= G= \exp \frak{g}$. We will also be interested in a second bundle 
 \begin{align}
 N &\cong T^*M \oplus \Lambda^4 T^*M \oplus( T^*M\otimes \Lambda^6 T^*M) \,. 
\end{align}
The action of sections of these bundles,
\begin{align}
  V  = v + \nu_2 + \nu_5\ \in \Gamma (E)\,, \quad W = w + \omega_2 + \omega_5\ \in \Gamma (E)\,, \quad  
 \mathcal{X} &= \chi_1 + \chi_4 + \chi_{1,6}\ \in \Gamma(N)\,,
\end{align}
is mediated by the generalised Lie derivative \cite{Berman:2011cg,Coimbra:2011ky,Berman:2012vc} defined as 
\begin{align}
  \gL_{V} W &= [v,w] + \bigl(L_v \omega_2 - \imath_w d\nu_2\bigr) + \bigl( L_v \omega_5 - \imath_w d\nu_5 - \omega_2 \wedge d\nu_2 \bigr) \,,
\label{eq:gLie}
\\
 \gL_V \mathcal{X} &= L_v \chi_1 + \bigl(L_v \chi_4 - \chi_1 \wedge d\nu_2\bigr)   + \bigl(L_v \chi_{1,6} + j\chi_4 \wedge d\nu_2 + j\chi_1\wedge d \nu_5\bigr) \,.
\end{align}
We define \cite{Coimbra:2011ky} a symmetric bilinear map  $\langle\cdot ,\,\cdot \rangle : E\times E\to N$ as 
\begin{align}
 \langle V,\,W \rangle 
 &= (\imath_v \omega_2+\imath_w \nu_2) + (\imath_v \omega_5 - \nu_2\wedge \omega_2 + \imath_w\nu_5)   + (j\nu_2\wedge \omega_5 + j\omega_2\wedge \nu_5)\,,
\end{align}
such that  the generalized Lie derivative satisfies
\begin{align}
 \langle \gL_U V,\,W\rangle + \langle V,\,\gL_U W\rangle = \gL_U \langle V,\,W \rangle \, \quad \forall U,V,W \in \Gamma(E).
\label{eq:gL-bracket}
\end{align}
  
The parallelisation consists of a set of sections $E_A \in \Gamma(E)$ that:
\begin{itemize}
    \item form a globally defined basis for $\Gamma(E)$
    \item give rise to an $E_{6(6)}$ element\footnote{An extension of this setup is to allow $E_{A}{}^M$ to be elements of $E_{6(6)}\times \mathbb{R}^+$, though for simplicity in the presentation we shall demand no $\mathbb{R}^+$ weighting.},  $E_{A}{}^M$, whose matrix entries are the components of $E_A$ 
    \item realise the algebra of the EDA through the generalised Lie derivative 
 \begin{equation}\label{eq:framealg}
 {\cal L}_{E_A} E_B =- X_{AB}{}^C E_C \, ,
 \end{equation}
 where the constants $X_{AB}{}^C$ are the same as those defined   through the relations in eq. \eqref{eq:TheEDA} and obey the Leibniz identity.
\end{itemize}
 
The parallelisation can be directly constructed in terms of the  right-invariant Maurer-Cartan one-forms on $G$, $r^a$, their dual vector fields $e_a$,  and the trivector, $\lambda^{a_1 a_2 a_3}$, and  hexavector, $\lambda^{a_1 \dots a_6}$. This can thus be thought of as a special example of the more general prescription of \cite{Inverso:2017lrz}, where we only make use of the aforementioned geometric data on the $(3,6)$-Nambu-Lie Group $G$. Following the decomposition of EDA generators we write $ {E}_{A}  = \{   E_{a},\, E^{a_1 a_2} ,\, E^{a_1 \dots a_5} \}$ with
\begin{align}
\begin{split}
 E_a &= e_a\,, \qquad
 E^{a_1a_2} = - \lambda^{a_1a_2b}\,e_b + r^{a_1} \wedge r^{a_2}\,,
\\
 E^{a_1\ldots a_5} &= \bigl( \lambda^{a_1\ldots a_5b}+ 5\,\lambda^{[a_1a_2a_3}\,\lambda^{a_4a_5]b}\bigr)\,e_b - 10\,\lambda^{[a_1a_2a_3}\,r^{a_4} \wedge r^{a_5]} + r^{a_1} \wedge \ldots\wedge r^{a_5} \,.
\end{split}
\label{eq:frames}
\end{align}
 It is straightforward, but indeed quite lengthy, to verify that these furnish the EDA algebra.  A first check is to see that after using  the identities \eqref{eq:dlambda} to evaluate derivatives we can go to the identity of $M$ where  $\lambda^{a_1 a_2 a_3}$ and $\lambda^{a_1 \dots a_6}$ vanish.  One then has to use the adjoint invariance conditions that follow from eq \eqref{eq:adinvariance} to conclude that this holds away from the identity.
 
 If we specialise now to the case of $f_{b}{}^{a_1\dots a_6}=0$, which we recall is enforced for $n\leq 6$ by the fundamental identities, we find quickly an immediate consequence of eq. \eqref{eq:adinvariance} is that $d\lambda^{a_1\dots a_6} =0 $, and since $\lambda^{a_1\dots a_6} $ vanishes at the identity, it must be identically zero.  The remaining  adjoint invariance conditions can be combined to imply that
 \begin{align}
 \begin{split}
 f_{ab}{}^c\, \lambda^{abd} =0 \, ,  \quad   f_{af}{}^{[b_1|}\,\lambda^{f|b_2b_3}\,\lambda^{b_4b_5b_6]} = 0\,, \quad f_a{}^{[b_1b_2b_3}\,\lambda^{b_4b_5b_6]} = 0 \, ,\\ 
  f_d{}^{b_1b_2c}\, \lambda^{a_1a_2 d}- 3\,f_d{}^{a_1a_2 [c}\, \lambda^{b_1b_2] d} - 3\, f_{de}{}^{[b_1}\, \lambda^{b_2 c]d}\, \lambda^{a_1a_2e} - 3\, f_{de}{}^{[a_1}\, \lambda^{a_2]d[b_1}\, \lambda^{b_2c] e} =0\,.
 \end{split}
 \end{align}
These conditions are sufficient to ensure that frame algebra is obeyed. 

We also define the generalized frame field $\cE_\cA$, which is a section of $N$, through
\begin{align}
 \langle E_A,\,E_B\rangle = \eta_{AB}{}^{\cC}\,\cE_\cC \,,
\label{eq:cE-def}
\end{align}
where $\eta_{AB}{}^{\cC}$ is an invariant tensor of the $E_{n(n)}$.    For $E_{6(6)}$  this tensor is related to the symmetric invariant (see the appendix for details) such that the  explicit form of $\cE_\cA$ has components   
 \begin{align}
 \begin{split}
  \cE^a &= r^a\,,\qquad \cE^{a_1\ldots a_4} =4\,\lambda^{[a_1a_2a_3}\,r^{a_4]} + r^{a_1\ldots a_4}    \,,
 \\
  \cE^{a',a_1\ldots a_6}&=  (\lambda^{a_1\ldots a_6}\,r^{a'}-30\,\lambda^{a'[a_1a_2}\,\lambda^{a_3a_4a_5}\,r^{a_6]}) -15\,\lambda^{a'[a_1a_2}\,r^{a_3\ldots a_6]} + jr^{a'}  r^{a_1\ldots a_6}   \,. 
 \end{split}
 \end{align}
 Here we denote $r^{a_1 \ldots a_m} = r^{a_1} \wedge \dots \wedge r^{a_m}$ and  make use of the $j$-wedge contraction  of \cite{Pacheco:2008ps} to deal with mixed symmetry fields.\footnote{For a $p+1$-form $\alpha$ and a $(n-p)$-form $\beta$, we define
\begin{equation}
\left( j\alpha \wedge \beta\right)_{i,i_1\ldots i_n} = \frac{n!}{p!(n-p)!} \alpha_{i[i_1 \ldots i_p} \beta_{i_{p+1} \ldots i_n]} \,.
\end{equation}} One can consider now the action of the frame field $E_A$ on these $\cE_{\cA}$ and by virtue of eq. \eqref{eq:cE-def}, again find that they furnish the EDA algebra, albeit in a different representation as described in the appendix.   
 
\subsection{Generalised Scherk-Schwarz reductions and the Embedding Tensor} 
The generalised frame field introduced above can be used as a compactification Ansatz within ExFT known as a generalised Scherk-Schwarz reduction.  In this procedure all internal coordinate dependence is factorised into dressings given by the generalised frame.  The algebra in eq. \eqref{eq:framealg} ensures the dimensional reduction results in a lower dimensional gauged supergravity.  The structure constants of the EDA determine the gauge group of this lower dimensional theory, and in such a context are known as the embedding tensor.  To facilitate contact with the literature \cite{LeDiffon:2008sh} we can express this  in terms of the $\overline{{\bf 27}}$ and ${\bf 351}$ representations of $E_{6(6)}$ as   
\begin{equation}
X_{AB}{}^C= d_{AB D}Z^{CD} + 10 d_{ADS} d_{BRT} d^{CDR} Z^{ST}  - \frac{3}{2} \vartheta_{[A} \delta_{B]}^C  - \frac{15}{2} d_{AB D} d^{CDE}\vartheta_E \, .
\end{equation} 
The components of the antisymmetric $Z^{AB}$ are determined to be 
\begin{align}
\begin{split}
 Z^{ab}&=0\,,\\  
 Z_{a_1a_2}{}^b &=-\tfrac{5 }{2 \cdot 4!\sqrt{10}}  \,f_d{}^{db c_1\ldots c_4 }\,\epsilon_{a_1a_2 c_1\ldots c_4  } \, (=0)  \,, 
\\
 Z_{a_1\ldots a_5 }{}^b &= - \tfrac{5}{2\sqrt{10}}  \,f_d{}^{dbc}\,\epsilon_{a_1\ldots a_5 c}  \,,
\\
 Z_{a_1 a_2\, , \, b_1 b_2} &= - \tfrac{5}{ 3! \sqrt{10}}   \,(f_{[a_1}{}^{c_1c_2c_3}\,\epsilon_{a_2]c_1c_2c_3 b_1 b_2 }- f_{[b_1}{}^{c_1c_2c_3}\,\epsilon_{b_2]c_1c_2c_3 a_1 a_2 } ) \,,
\\
 Z_{a_1 a_2 \, , \, b_1 \ldots b_5} &= - \tfrac{5}{2\sqrt{10}}   \,(f_{a_1 a_2 }{}^c\,\epsilon_{c b_1 \ldots b_5}- 10 f_{[b_1 b_2}{}^c\,\epsilon_{b_3 b_4 b_5] c a_1 a_2}) \,, 
\\
 Z_{a_1\dots a_5 \, , \, b_1 \dots b_5 } &= 0\,,
\end{split}
\end{align}
and those of $\vartheta_A$ (sometimes called the trombone gauging) to be 
\begin{align}
 \vartheta_a{}  = \frac{f_{ac}{}^c}{3} \,, \quad
\vartheta^{a_1 a_2 } = - \frac{f_c{}^{ca_1 a_2}}{3} \,,\quad
\vartheta^{a_1\ldots a_5 } = - \frac{f_c{}^{ca_1\ldots a_5}}{3} \, (= 0)  \, . 
\end{align} 

\section{Yang-Baxter-ology} 
 \label{s:YB}
 \subsection{EDA via $\rho$-twisting}
In the context of DFT, Yang-Baxter deformations can be understood as the $\Odd$ transformation, generated by a bivector, acting on a Drinfel'd double with vanishing dual structure constants \cite{Araujo:2017jkb,Sakamoto:2017cpu,Sakamoto:2018krs,Bakhmatov:2018apn,Bakhmatov:2018bvp,Catal-Ozer:2019tmm}. The bivector that generates the transformation is then related to the classical $r$-matrix, the dual structure constants are coboundaries and the requirement that the $\Odd$ transformed algebra is a Drinfel'd double is precisely the classical Yang-Baxter equation.

This suggests a natural generalisation of Yang-Baxter deformations to EDAs \cite{Sakatani:2019zrs,Malek:2019xrf}. We begin with an EDA $\widehat{\frak{d}}_6$ with only the structure constants, $f_{ab}{}^c$, corresponding to a maximally isotropic Lie subalgebra $\frak{g}$, non-vanishing and $f_a{}^{bcd}=f_a{}^{b_1\ldots b_6}=0$, i.e.
\begin{equation}
    \begin{split}
        \hat{T}_a \circ \hat{T}_b &= f_{ab}{}^c\,\hat{T}_c \,,\\
        \hat{T}_a \circ \hat{T}^{b_1b_2} &= 2\,f_{ac}{}^{[b_1}\,\hat{T}^{b_2]c}\,,\\
        \hat{T}_a \circ \hat{T}^{b_1\ldots b_5} &= - 5\,f_{ac}{}^{[b_1}\,\hat{T}^{b_2\ldots b_5]c} \,,\\
        \hat{T}^{a_1a_2} \circ \hat{T}_b &= 3\,f_{[c_1c_2}{}^{[a_1}\,\delta^{a_2]}_{b]}\,\hat{T}^{c_1c_2}\,,\\
        \hat{T}^{a_1a_2} \circ \hat{T}^{b_1b_2} &= f_{c_1c_2}{}^{[a_1}\,\hat{T}^{a_2]b_1b_2c_1c_2}\,,\\
        \hat{T}^{a_1a_2} \circ \hat{T}^{b_1\ldots b_5} &= 0 \,,\\
        \hat{T}^{a_1\ldots a_5} \circ \hat{T}_b &= 5\,f_{bc}{}^{[a_1}\,\hat{T}^{a_2\ldots a_5]c} + 10\,f_{c_1c_2}{}^{[a_1}\,\delta^{a_2}_b\,\hat{T}^{a_3a_4a_5]c_1c_2} \,,\\
        \hat{T}^{a_1\ldots a_5} \circ \hat{T}^{b_1b_2} &= 0\,,\\
        \hat{T}^{a_1\ldots a_5} \circ \hat{T}^{b_1\ldots b_5} &=0 \,.
    \end{split}
\end{equation}
We denote the structure constants collectively as $\hat{X}_{AB}{}^C$\,. 

We now perform an $E_{6(6)}$ transformation of the above EDA by a trivector, $\rho^{abc}$, and hexavector, $\rho^{a_1\ldots a_6}$, which will play the analogue of the classical $r$-matrix.  The corresponding $E_{6(6)}$ group element is given by
\begin{align} \label{eq:YBtr}
 C_A{}^B\equiv \bigl(e^{\frac{1}{6!}\,\rho^{a_1\ldots a_6}\,R_{a_1\ldots  a_6}}e^{\frac{1}{3!}\,\rho^{a_1a_2a_3}\,R_{a_1a_2a_3}}\bigr)_A{}^B\,,
\end{align}
in which the generators $R_{a_1a_2 a_3}$ and $R_{a_1\ldots a_6}$ are specified in the appendix. Explicitly we have that  
\begin{align}
 (C_A{}^B) = \begin{pmatrix}
 \delta_a^b & 0  & 0 \\
 \frac{\rho^{b a_1a_2}}{\sqrt{2!}} & \delta^{a_1a_2}_{b_1b_2} & 0 \\
 \frac{\tilde{\rho}^{b;a_1\cdots a_5}}{\sqrt{5!}} & \frac{20\,\delta_{b_1b_2}^{[a_1a_2} \rho^{a_3a_4a_5]}}{\sqrt{2!\,5!}} & \delta^{a_1\cdots a_5}_{b_1\cdots b_5} \end{pmatrix} \, , 
\end{align} 
where 
\begin{align}
\tilde{\rho}^{b;a_1\ldots a_5} &\equiv \rho^{ba_1\ldots a_5}+5\,\rho^{b[a_1a_2}\,\rho^{a_3a_4a_5]} \,.
\end{align} 
Equivalently, we twist the generators by the group element \eqref{eq:YBtr} resulting in
\begin{equation}
  \begin{split}
T_a &= \hat{T}_a\,,\qquad 
T^{a_1a_2} = \hat{T}^{a_1a_2} + \rho^{ba_1a_2}\,\hat{T}_b\,,
\\
T^{a_1\ldots a_5} &= \hat{T}^{a_1\ldots a_5} + 10\,\rho^{[a_1a_2a_3}\,\hat{T}^{a_4a_5]} + \tilde{\rho}^{b;a_1\ldots a_5} \,\hat{T}_b\,.
   \end{split}
\end{equation}

For the twisted generators, we obtain $T_A \circ T_B =X_{AB}{}^C\,T_C$ with
\begin{align}
 X_{AB}{}^C \equiv C_A{}^D \,C_B{}^E\,(C^{-1})_F{}^C\,\hat{X}_{DE}{}^F\,.
\label{eq:cF-def}
\end{align}

We now require that the new algebra defines an EDA $\frak{d}_6$. This imposes conditions on $\rho^{abc}$ and $\rho^{a_1\ldots a_6}$ and we will interpret these as analogues of the classical Yang-Baxter equation. From the products $T_a\circ T^{b_1b_2}$  and $T_a\circ T^{b_1\ldots b_5}$\,, the dual structure constants are identified as
\begin{align}
f_a{}^{b_1b_2b_3} = 3\,f_{ac}{}^{[b_1}\,\rho^{|c|b_2b_3]}\,, \qquad
f_a{}^{b_1\ldots b_6} = 6\,f_{ac}{}^{[b_1|}\,\tilde{\rho}^{c;|b_2\ldots b_6]} (=0)\,,
\end{align}
which take the form of (twisted) coboundaries \eqref{eq:coboundary} and the $(=0)$ holds for $\frak{d}_6$. The remaining products impose a number of conditions, of which the following is particularly intriguing
\begin{equation}
    \begin{split} \label{eq:YB}
         3\, f_{d_1d_2}{}^{[b_1}\,\rho^{b_2b_3]d_1}\,\rho^{a_1a_2d_2} &= f_{d_1d_2}{}^{[a_1}\,\tilde{\rho}^{a_2];b_1b_2b_3 d_1d_2} \,.
    \end{split}
\end{equation} 

This is a natural generalisation of the classical Yang-Baxter (YB) equation which we will elaborate more on later. In general, we get a further set of conditions which are required to ensure that the new algebra defines an EDA $\frak{d}_6$. Mostly these additional conditions appear rather cumbersome but we note the requirement  that 
\begin{equation}
    \begin{split} \label{eq:YB-2}
      \rho^{a_1 a_2 b}f_{a_1 a_2}{}^c &=0\,. \\
    \end{split}
\end{equation}
With $f_a{}^{b_1\ldots b_6}=0$, the Bianchi identity for $f_{ab}{}^c$ together with the generalised Yang-Baxter equation  eq. \eqref{eq:YB} and compatibility condition eq. \eqref{eq:YB-2} imply the fundamental identity for $f_{a}{}^{b_1\dots b_3}$.  Indeed  since the Leibniz identity \eqref{eq:Leibniz} is $E_{6(6)}$-invariant, it is guaranteed to hold for $X_{AB}{}^C$. Therefore, we see that the generalised Yang-Baxter equation \eqref{eq:YB} together with the other conditions obtained by imposing that the new algebra is an EDA imply that the new dual structure constants satisfy their fundamental identities \eqref{eq:FI-3} and \eqref{eq:FI-6} and the condition \eqref{eq:addcon}.  
 
In \cite{Bakhmatov:2019dow}, a different approach was taken using a generalisation of the open/closed string map to propose a generalisation of the classical YB equation for a trivector deformation of 11-dimensional supergravity. The approach of \cite{Bakhmatov:2019dow} is not limited to group manifolds, unlike the present case, but also only considers trivector deformations. However, when specialising \cite{Bakhmatov:2019dow} to group manifolds and considering our deformations with $\rho^{a_1\ldots a_6} = 0$, the resulting equation of \cite{Bakhmatov:2019dow} is different and, in particular, weaker than the YB equation we find here \eqref{eq:YB} with $\rho^{a_1\ldots a_6} = 0$, or indeed the $\SL{5}$ case discussed in \cite{Sakatani:2019zrs,Malek:2019xrf}. Indeed, as shown in \cite{Bakhmatov:2020kul} based on explicit examples, the proposed YB-like equation of \cite{Bakhmatov:2019dow} is not sufficient to guarantee a solution of the equations of motion of 11-dimensional supergravity, while our deformations subject to the above conditions preserve the equations of motion of 11-dimensional supergravity by construction.

\subsection{ Nambu 3- and 6-brackets from $\rho$-twisting} \label{s:YBBrackets}
The trivector $\rho^{a_1a_2a_3}$ and hexavector $\rho^{a_1\ldots a_6}$ define 3- and 6-brackets via \eqref{eq:coboundary}. First define the maps
\begin{equation}
    \rho_3: \frak{g}^* \wedge \frak{g}^* \longrightarrow \frak{g} \,, \qquad \tilde{\rho}_6: \Lambda^5 \frak{g}^* \longrightarrow \frak{g} \,,
\end{equation}
as
\begin{equation}
    \begin{split}
    \rho_3(x_1,x_2) &= \rho^{abc}\, (x_1)_b\, (x_2)_c\, T_a \,, \\
    \tilde{\rho}_6(x_1,\ldots, x_5) &= \tilde{\rho}^{a_1; a_2 \ldots a_6}\, (x_1)_{a_2} \ldots (x_5)_{a_6}\, T_{a_1} \,, \; \forall\; x_1,\, \ldots,\, x_5 \in \frak{g}^* \,.
    \end{split}
\end{equation}
This allows the generalised Yang-Baxter equation eq. \eqref{eq:YB} to be cast in a basis independent way
\begin{equation}
    \begin{split} \label{eq:YB-bi}
        x_1 \left( \left( ad_{\rho_3(y_1,y_2)} \rho_3 \right)(x_2,x_3) \right) + y_1 \left( \tilde{\rho}_6(\Delta(y_2),x_1,x_2,x_3) \right) - y_2 \left( \tilde{\rho}_6(\Delta(y_1),x_1,x_2,x_3) \right) &= 0 \,,
    \end{split}
\end{equation}
for all  $y_1,\,y_2,\,x_1,\,x_2,\,x_3 \in \frak{g}^*$. Note that the first term in \eqref{eq:YB-bi} is automatically antisymmetric in $\left(x_1,\,x_2,\,x_3\right)$ due to antisymmetry of $\rho^{a_1a_2a_3}$.

Then, the associated 3- and 6-brackets are defined as
\begin{equation}
    \begin{split} \label{eq:YBbrackets}
        \left\{ x_1,\,x_2,\,x_3\right\}_3 &= ad_{\rho_3(x_1,x_2)} x_3 + ad_{\rho_3(x_2,x_3)} x_1 + ad_{\rho_3(x_3,x_1)} x_2 \,, \\
        \left\{ x_1,\, \ldots,\, x_6 \right\}_6 &= ad_{\tilde{\rho}_6(x_{1},\ldots, x_{5})} x_{6} + \text{cyclic permutations} \,,
    \end{split}
\end{equation}
for all $x_1,\, \ldots ,\, x_6 \in \frak{g}^*$.
Alternatively, to match more with the usual discussion of classical $r$-matrices in integrability, we can use the Cartan-Killing form on $\frak{g}$ to define 3- and 6-brackets on $\frak{g}$. For this, it is more convenient to define $\rho'_3$ and $\rho'_6$ as
\begin{equation} 
        \rho'_3: \frak{g} \wedge \frak{g} \longrightarrow \frak{g} \,, \qquad \tilde{\rho}'_6: \Lambda^5 \frak{g} \longrightarrow \frak{g} \,, 
\end{equation}
with
\begin{equation}
    \rho'_3(x_1,x_2) = \rho_3(\kappa(x_1),\kappa(x_2)) \,, \qquad \tilde{\rho}'_6(x_1,\ldots,x_6) = \tilde{\rho}_6(\kappa(x_1),\ldots,\kappa(x_5)) \,,
\end{equation}
and where $\kappa$ is the Cartan-Killing metric viewed as a map $\kappa: \mathfrak{g} \longrightarrow \mathfrak{g}^*$. Now, the 3- and 6-brackets on $\frak{g}$ are defined as
\begin{equation}
    \begin{split}\label{eq:YBbrackets2}
        \left\{x_1,\,x_2,\,x_3\right\}_3 &= \left[ x_1, \rho'_3(x_2,x_3) \right] + \left[ x_2,\rho'_3(x_3,x_1) \right] + \left[ x_3,\rho'_3(x_1,x_2) \right] \,, \\
        \left\{ x_1,\, \ldots,\, x_6 \right\}_6 &= \left[ x_1, \tilde{\rho}'_6(x_{2},\ldots,x_6) \right] + \text{cyclic permutations} \,.
    \end{split}
\end{equation}

The generalised Yang-Baxter equation \eqref{eq:YB} together with the other constraints required such that the new algebra is an EDA, such as \eqref{eq:YB-2}, imply that the 3- and 6-brackets defined above in \eqref{eq:YBbrackets} and \eqref{eq:YBbrackets2} satisfy their fundamental identities \eqref{eq:FI-3} and \eqref{eq:FI-6}.

 \subsection{The generalised YB equation}
 To understand better the generalised YB equation obtained above, let us adopt a tensor product notation $\rho_{124} = \rho^{abc} T_{a}\otimes T_{b} \otimes 1 \otimes T_c \otimes 1 $ etc. such that the indices denote the contracted slots in a tensor product of $\frak{g}$.  Assuming that \eqref{eq:YB-2} holds and that $\rho^{a_1\ldots a_6}= 0$ we have that \eqref{eq:YB} becomes
\begin{align}\label{eq:YBtensor}
 &[\rho_{123},\rho_{145}]
 + [\rho_{123},\rho_{245}]
 + [\rho_{123},\rho_{345}]
\nonumber\\
 &+ \frac{1}{2}\bigl(
 [\rho_{124}+\rho_{125},\rho_{345}]
 +[\rho_{234}+\rho_{235},\rho_{145}]
 +[\rho_{314}+\rho_{315},\rho_{245}] \bigr) = 0\,.
\end{align}  
Introducing a (anti-)symmetrizer in the tensor product $\sigma_{[123],[45]} $, allows this equation to be concisely given as
\begin{equation}\label{eq:YBtensor2}
\sigma_{[123],[45]}  [\rho_{123} +\rho_{234} , \rho_{145}]   = 0 \, . 
\end{equation}  

 Suppose that we have a preferred $q\in\frak{g}$ such that
\begin{equation}\label{eq:rhotor}
    \rho_{123} = r_{ 12} \otimes q_{3 } + r_{ 23} \otimes q_{1}  - r_{ 13} \otimes q_{2}  \,,\quad r_{12} = \sum_{a,b\neq q} r^{ab}T_{a}\otimes T_b  = - r_{21}\, , 
\end{equation}
with $r_{12}$ neutral (i.e. $[r_{12},q_1]= 0$)  
then we find eq. \eqref{eq:YBtensor} becomes 
\begin{align}
 {\tt  YB}_{[12|4|} \otimes  q_{3]} \otimes  q_{5} -    {\tt   YB}_{[12|5|} \otimes q_{3]} \otimes  q_{4}  =  0\, , 
\end{align}
in which
\begin{equation}\label{eq:cYB}
 {\tt  YB}_{123} = [r_{12}, r_{13}] + [r_{12}, r_{23}] + [r_{13}, r_{23}]\, ,
\end{equation} 
is the classical Yang-Baxter equation for $r$. 

Recall that the classical YB equation arises from the quantum one
\begin{equation}
    \mathbb{R}_{13}\mathbb{R}_{12}\mathbb{R}_{32} = \mathbb{R}_{32}\mathbb{R}_{12}\mathbb{R}_{13}
\end{equation}
as the leading terms in the `classical' expansion $\mathbb{R}_{12} = 1 + \hbar\,r_{12} + O(\hbar^2)$.  An obvious question is if there is an equivalent `quantum' version of eq. \eqref{eq:YBtensor2}?  We give here one proposal (with no claim of first principle derivation or uniqueness) for such a starting point.   Let us define\footnote{We use lower case Roman indices to denote tensor product locations.} in the semi-classical limit 
\begin{align}\label{eq:Rsemicl}
\mathbb{R}_{i;jk} &= 1 + \hbar\, \rho_{ijk} + O(\hbar^2) \, , \\
\mathbb{R}_{ij;kl} &= 1 + \frac{\hbar}{4} \left( \rho_{ijk}+\rho_{ijl} -  \rho_{ikl}-\rho_{jkl} \right)+ O(\hbar^2)\, . 
\end{align}
Then eq. \eqref{eq:YBtensor} follows from 
\begin{equation}\label{eq:Rguess}
    \sigma_{[123],[45]} \mathbb{R}_{1;23} \mathbb{R}_{23;45} \mathbb{R}_{1;45}  =  \sigma_{[123],[45]}  \mathbb{R}_{1;45} \mathbb{R}_{23;45} \mathbb{R}_{1;23} \, . 
\end{equation}
 This view point is very suggestive that this may just represent a standard Yang-Baxter equation for the scattering of $\wedge^2\frak{g},\wedge^2\frak{g}$ and $\frak{g}$ obtained by S-matrix fusion.   Here we leave an exploration of this as an open direction; further work is required to understand which quantum R-matrices give rise under S-matrix fusion to an $\mathbb{R}_{i;jk}$ and $\mathbb{R}_{ij;kl}$ with the expansion  \eqref{eq:Rsemicl} and what are the resultant $\rho_{ijk}$. Conversely one might ask if there exist solutions of   \eqref{eq:Rguess} compatible \eqref{eq:Rsemicl} but that are not obtained from fusion? 
 
\begin{figure}[tbp!]
\begin{center}
        \includegraphics[width=0.9\textwidth]{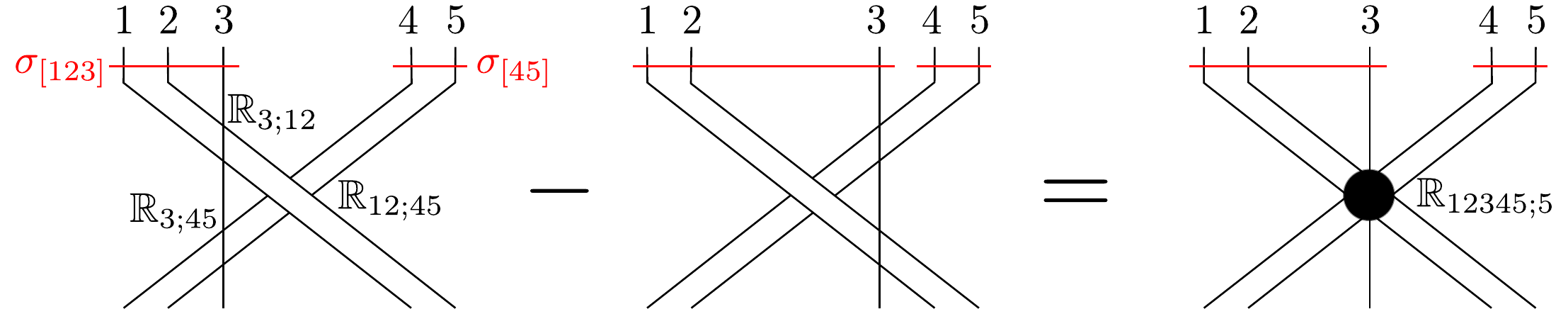}
\end{center}
\vskip -0.6cm
       \caption{  A proposed schematic for the generalised Yang-Baxter equation. The red lines indicate anti-symmetrisation and the black circle is a contact term that gives rise in the semi-classical limit to a contribution involving $\rho_6$. }
    \label{fig:fusion}
\end{figure}

Restoring $\rho^{a_1\ldots a_6}$   we can amend this equation to 
\begin{equation}
\sigma_{[123],[45]}  [\rho_{123} +\rho_{234} , \rho_{145}]   =   \frac{1}{2}\left( \rho_{12345;5} +\rho_{12345;4}  \right)\,, 
\end{equation}
where, for example,
\begin{equation}
      \rho_{12345;5} \equiv \rho^{abcdef}\,T_a\otimes T_b \otimes T_c\otimes T_d \otimes [T_e,T_f] \,.
\end{equation}
This is somewhat suggestive of a contact term in the YB relation that may lead to a quantum version of the form
 \begin{equation}\label{eq:Rguesswithrho6}
    \sigma_{[123],[45]} \mathbb{R}_{1;23} \mathbb{R}_{23;45} \mathbb{R}_{1;45}  - \sigma_{[123],[45]}  \mathbb{R}_{1;45} \mathbb{R}_{23;45} \mathbb{R}_{1;23} =  \sigma_{[123],[45]} \mathbb{R}_{12345;5}   \, . 
\end{equation}
 Pictographically this is indicated in figure \ref{fig:fusion}. 
\section{Examples} \label{s:Examples}

In this section we wish to present a range of examples of the EDA, both of coboundary type and otherwise.  We will give some broad general classes that correspond to embedding the algebraic structure underlying existing T-dualities of the type II theory.  In addition, in the absence of a complete classification, here we provide   a selection of specific examples.   

\subsection{Abelian}

When the subalgebra $\{T_a\}$ is Abelian, arbitrary $\rho^{a_1 a_2 a_3 }$ and $\rho^{a_1 \ldots a_6}$ are solutions of Yang--Baxter-like equations.  However $f_a{}^{b_1b_2 b_3}=0$ and $f_a{}^{b_1 \ldots b_6}=0$ and the EDA is Abelian. 

\subsection{Semi-Abelian EDAs and Three Algebras} 

The algebraic structure corresponding to non-Abelian T-duality is a semi-abelian Drinfel'd double i.e. a double constructed from  some $n-1$ dimensional Lie-algebra (representing the non-Abelian isometry group of the target space) together with a $U(1)^{n-1}$ (or perhaps $\mathbb{R}_+^{n-1}$) factor.  An analogue here would be  to take $f_{ab}{}^c \neq 0 $ and $f_{a}{}^{b_1\dots b_3}=0$, this however is not especially interesting.  More intriguing is to consider the analogue of the picture {\em after} non-Abelian T-dualisation has been performed in which   the $U(1)^{n-1}$ would be viewed as the physical space.  This motivates the case of semi-Abelian EDAs with $f_{ab}{}^c = 0 $ but $f_a{}^{b_1 \dots b_3} \neq 0$.   

In this case  the Leibniz identities reduce to the fundamental identities
\begin{align}
 f_a{}^{d c_1c_2}\, f_d{}^{b_1b_2b_3} - 3\, f_d{}^{c_1c_2[b_1}\,f_a{}^{b_2b_3]d} =0\,,\qquad f_a{}^{b_1\cdots b_6}=0\,.
\end{align}
Each solution for this identity gives an EDA. To identify these one can use existing classification efforts  and considerations  of three algebras that followed in light of their usage \cite{Bagger:2006sk} to describe theories of interacting multiple M2 branes. 

The first case to consider are the Euclidean three algebras, such that $f^{b_1\dots b_4}= f_{a}{}^{b_1\dots b_3} \delta^{ab_4} $ is totally antisymmetric.  Here the fundamental identity is very restrictive and results in a unique possibility: the  four-dimensional Euclidean three algebra\cite{Nagy:2007cle,Papadopoulos:2008sk,Gauntlett:2008uf}, whose structure constants are just the antisymmetric symbol, complemented with two $U(1)$ directions.  Relaxing the requirement of a positive definite invariant inner product allows a wider variety \cite{Gomis:2008uv,Benvenuti:2008bt,Ho:2008ei,deMedeiros:2008bf,Ho:2009nk,deMedeiros:2009hf}.  Dispensing the requirement of an invariant inner product (which thus far appears unimportant for  the EDA) allows non-metric three algebras \cite{Awata:1999dz, Gustavsson:2008dy,Gran:2008vi}.\footnote{In addition there are three algebra structures \cite{Bagger:2008se,Chen:2009cwa} in which $f_d{}^{abc}$ is not totally antisymmetric in its upper indices.  These can be used to describe interacting 3d theories with lower supersymmetry.  It is a unclear if they could play role in the context of EDAs.}

\subsection{$r$-matrix EDAs} 
We now consider coboundary EDAs given in terms of an $r$-matrix as in eq.\eqref{eq:rhotor} obeying the YB equation \eqref{eq:cYB}.  Splitting the generators of $\frak{g}$ into $T_{\bar{a}}$ with $\bar{a} = 1,\dots, 5$ and $T_6$ (identified with the generator $q$ appearing in \eqref{eq:rhotor}) we have the non-vanishing components $\rho^{\bar{a}\bar{b} 6} = r^{\bar{a}\bar{b}}$.  Furthermore the condition \eqref{eq:YB-2} requires that 
\begin{equation} \label{eq:rf}
  r^{\bar{a}\bar{b}}  f_{\bar{a} \bar{b}}{}^6  =   r^{\bar{a}\bar{b}}  f_{\bar{a} \bar{b}}{}^{\bar{c}} = r^{\bar{a}\bar{b}}  f_{\bar{a} 6 }{}^{\bar{c}} =r^{\bar{a}\bar{b}}  f_{\bar{a} 6 }{}^6 =0 \, ,
\end{equation}
in which the last two equalities match the statement that $r$ is neutral under $T_6$.  In such a  setup,   the   dual structure constants are specified as 
\begin{align}
\begin{split} 
    f_{\bar{a}}{}^{\bar{b}_1\bar{b}_2 6} = 2 f_{\bar{a}\bar{c}}{}^{[\bar{b}_1} r^{|\bar{c}|\bar{b}_2]} + f_{\bar{a}6}{}^6 r^{\bar{b}_1\bar{b}_2} \, , \quad f_{\bar{a}}{}^{\bar{b}_1\bar{b}_2\bar{b}_3} = 3 f_{\bar{a}6}{}^{[\bar{b}_1}r^{\bar{b}_2 \bar{b}_3]}\,  ,  \quad  f_{6}{}^{\bar{b}_1\bar{b}_2 6} = f_{6}{}^{\bar{b}_1\bar{b}_2 \bar{b}_3} = 0 \, .  
\end{split}
\end{align} 
Assuming further that $\bar{\frak{g}} =\textrm{span}(T_{\bar{a}}) $ is a sub-algebra of $\frak{g}$ then $r^{\bar{a}\bar{b}}$ defines an $r$-matrix on $\bar{\frak{g}}$ obeying the YB equation. Consequently  $\tilde{f}^{\bar{b}_1\bar{b}_2}{}_{\bar{a} }=- 2 f_{\bar{a}\bar{c}}{}^{[\bar{b}_1} r^{|\bar{c}|\bar{b}_2]}  $ are the structure constants of a dual Lie algebra $\bar{\frak{g}}_R$ and $\bar{\frak{d}} = \bar{\frak{g}} \oplus \bar{\frak{g}}_R $ is a Drinfel'd double. Thus we have a family of embeddings of the Drinfel'd double into the EDA specified by $f_{\bar{a}6}{}^6$   and $f_{\bar{a}6}{}^{\bar{b}}$.   When $\frak{g} = \bar{\frak{g}} \oplus u(1)$ is a direct sum (such that $f_{\bar{a}6}{}^6=f_{\bar{a}6}{}^{\bar{b}} = 0 $), then this is precisely an example of the non-metric three algebra of  \cite{Awata:1999dz, Gustavsson:2008dy,Gran:2008vi}. We emphasise though that not every (coboundary) double can be embedded in this way; one must still ensure that equation \eqref{eq:rf} holds. 

\subsection{Explicit examples} 
We now present a selection of explicit examples that illustrate coboundary and non-coboundary EDAs. 

\subsubsection{Trivial non-examples based on $SO(p,q)$}
To illustrate that the EDA requirements are indeed quite restrictive we can first consider the case of $\frak{g} =\frak{so}(p,q)$ with $p+q=4$.   A direct consideration of the Leibniz identities reveals that there is no non-zero solution for $f_a{}^{b_1\dots b_3}$ (in fact the cocycle conditions alone determine this).  Equally the Leibniz identities admit only trivial solutions in the case of  $\frak{iso}(p,q)= \frak{so}(p,q) \ltimes \mathbb{R}_{+}^{p+q}$ with $p+q =3$.

\subsubsection{An example both   coboundary and non-coboundary solutions} 
 
We consider an indecomposable nilpotent Lie algebra $N_{6,22}$ of \cite{doi:10.1063/1.522992} specified by the structure constants\footnote{Here we introduced a parameter $c_0$ for convenience, which is 1 in \cite{doi:10.1063/1.522992}.}
\begin{align}
 f_{12}{}^3 = 1\,,\quad
 f_{13}{}^5 = 1\,,\quad
 f_{15}{}^6 = c_0\,,\quad
 f_{23}{}^4 = 1\,,\quad
 f_{24}{}^5 = 1\,,\quad
 f_{34}{}^6 = c_0\,.
\end{align}

We find a family of solutions 
\begin{align}
\begin{split}
 f_1{}^{356} &= d_1\,,\quad
 f_1{}^{456} = d_2\,,\quad
 f_2{}^{156} = d_3\,,\quad
 f_2{}^{346} = -d_3\,,
\\
 f_2{}^{356} &= d_4\,,\quad
 f_2{}^{456} = d_5\,,\quad
 f_3{}^{456} = -d_1 + d_3\,,
\end{split}
\end{align}
which indeed satisfies the closure constraints. 
In particular, if we choose $c_0=0$, we can clearly see that this EDA contains a 10D Drinfel'd double $\{T_{\bar{a}},T^{\bar{a}6}\}$ ($\bar{a}=1,\dotsc,5$) as a Lie subalgebra.

We can also find $\rho^{abc}$ by considering the coboundary Ansatz. 
Supposing $c_0\neq 0$, the general solution to the generalised Yang-Baxter equation and compatibility condition is
\begin{align}
 \rho^{146} = e_1\,,\quad
 \rho^{156} = e_2\,,\quad
 \rho^{256} = e_3\,,\quad
 \rho^{346} = -e_2\,,\quad
 \rho^{356} = e_4\,,\quad
 \rho^{456} = e_5\,, 
\end{align}
where $e_1=0$ or $e_3=0$\,. 
The corresponding structure constants are
\begin{align}
\begin{split}
 f_1{}^{356}&=e_3\,,\quad
 f_1{}^{456}=e_2\,,\quad
 f_2{}^{156}=e_1\,,\quad
 f_2{}^{346}=-e_1\,,
\\
 f_2{}^{356}&=-2 e_2\,,\quad
 f_2{}^{456}=e_4\,,\quad
 f_3{}^{456}=e_1-e_3\,.
\end{split}
\end{align}
This means that only when $d_i$ ($i=1,\dotsc,5$) have the form
\begin{align}
 d_1=e_3\,,\quad
 d_2=e_2\,,\quad
 d_3=e_1\,,\quad
 d_4=-2e_2\,,\quad
 d_5=e_4 \,,
\end{align}
and satisfy $d_1=0$ or $d_3=0$, the cocycle becomes the coboundary.

\subsubsection{An example with $\rho^6$}

In the previous example  $\rho^6$ is absent. By considering the Lie algebra of the form $\mathfrak{g} =\mathfrak{g}_4\oplus u(1)\oplus u(1)$, where $\mathfrak{g}_4$ denotes a real 4D Lie algebra that is classified in \cite{Popovych:2003xb}, one can construct a number of examples\footnote{In the notation of  \cite{Popovych:2003xb} examples with $\rho^6\neq 0$ are found when $\mathfrak{g}_4$ is one of the following: $A_{3,1}+\mathfrak{u}(1)$,  $A_{3,4}^{-1}+\mathfrak{u}(1)$, $A_{3,5}^{0}+\mathfrak{u}(1)$, $A_{4,1}$, $A_{4,2}^{-2}$, $A_{4,5}^{a,b,-(a+b)}$,  $A_{4,6}^{-2b,b}$} (based on unimodular Lie algebras) that admit $\rho^6$.   To illustrate this let us consider the case that  $\mathfrak{g}_4=A_{4,1}$ specified by structure constants  \begin{align}
 f_{24}{}^1 = 1\,,\quad
 f_{34}{}^2  = 1\, . 
\end{align}
We find the generalised Yang-Baxter and compatibility equations admit the following family of solutions:
\begin{align}
\begin{split}
 \rho^{123} &= d_1\,,\quad
 \rho^{125} = d_2\,,\quad
 \rho^{126} = d_3\,,\quad
 \rho^{135} = \frac{d_1 d_4 d_8}{2 d_0}\,,\quad
 \rho^{136} = \frac{d_1 d_5 d_8}{2 d_0}\,,\quad
 \rho^{145} = d_4\,,
\\
 \rho^{146} &= d_5\,,\quad
 \rho^{156} = d_6\,,\quad
 \rho^{256} = d_7\,,\quad
 \rho^{356} = d_8\,,\quad
 \rho^{456} = \frac{2 d_0}{d_1}\,,\quad
 \rho^{123456} = d_0\,.
\end{split}
\end{align}
The corresponding dual structure constants are
\begin{align}
\begin{split}
 f_2{}^{156}&=\frac{2 d_0}{d_1}\,,\quad
 f_3{}^{125}=d_4\,,\quad
 f_3{}^{126}=d_5\,,\quad
 f_3{}^{256}=\frac{2 d_0}{d_1}\,,
\\
 f_4{}^{125}&=-\frac{d_1 d_4 d_8}{2 d_0}\,,\quad
 f_4{}^{126}=-\frac{d_1 d_5 d_8}{2 d_0}\,,\quad
 f_4{}^{156}=-d_7\,,\quad
 f_4{}^{256}=-d_8\,.
\end{split}
\end{align}  

\subsubsection{An $r$-matrix EDA}

In order to find a non-trivial example of the $r$-matrix EDAs, we consider a solvable Lie algebra $N_{6,29}^{\alpha\beta}$ of \cite{doi:10.1063/1.528721} defined by structure constants:
\begin{align}
 f_{13}{}^3 = 1\,,\quad
 f_{15}{}^5 = \alpha\,,\quad
 f_{16}{}^6 = 1\,,\quad
 f_{23}{}^3 = 1\,,\quad
 f_{24}{}^4 = 1\,,\quad
 f_{25}{}^5 = \beta\,,\quad
 f_{46}{}^3 = -1\,,
\end{align}
where $\alpha^2+\beta^2\neq 0$\,. 
This algebra contains a subalgebra generated by $\bar{\frak{g}} =\textrm{span}(T_{\bar{a}})$ ($\bar{a}=1,\dotsc,5$) and is non-trivial in the sense that it satisfies $f_{\bar{a}6}{}^{\bar{b}}\neq 0$ and $f_{\bar{a}6}{}^6\neq 0$. 

Supposing $\alpha\beta\neq 0$, we find the general solution for $\rho^{abc}$ is given by
\begin{align}
 \rho^{135} = c_1\,,\quad
 \rho^{235} = -c_1\,,\quad
 \rho^{356} = c_2\,,\quad
 \rho^{345} = c_3\,,
\end{align}
where $c_1=0$ when $\alpha\neq \beta$\,. 
The corresponding dual structure constants are
\begin{align}
\begin{split}
 f_1{}^{135}&=(1+\alpha) c_1\,,\quad
 f_1{}^{235}=-(1+\alpha) c_1\,,\quad
 f_1{}^{345}=(1+\alpha) c_3\,,\quad
 f_1{}^{356}=(2+\alpha) c_2\,,
\\
 f_2{}^{135}&=(1+\beta) c_1\,,\quad
 f_2{}^{235}=-(1+\beta) c_1\,,\quad
 f_2{}^{345}=(2+\beta) c_3\,,\quad
 f_2{}^{356}=(1+\beta) c_2\,,
\\
 f_4{}^{345}&=-c_1\,,\quad
 f_6{}^{356}=-c_1\,.
\end{split}
\end{align}
This solution contains an $r$-matrix EDA as a particular case $c_1=c_3=0$\,.

\section{Conclusion and Outlook}
In this work we have consolidated the exploration of exceptional Drinfel'd algebras introduced in \cite{Sakatani:2019zrs,Malek:2019xrf} extending the construction to the context of the $E_{6(6)} $ exceptional group. The algebraic construction here requires the introduction of a new feature: we have to consider not only a Lie algebra $\frak{g}$ together with a three-algebra specified by $f_3 \equiv f_a{}^{b_1\ldots b_3}$ as in \cite{Sakatani:2019zrs,Malek:2019xrf}, but we have to also include a six-algebra $f_6 \equiv f_a{}^{b_1\ldots b_6}$.  The Leibniz identities that the EDA must obey enforce  a set of  fundamental (Jacobi-like)  identities for the three- and six-algebra as well as some compatibility conditions. These compatibility conditions require that   $f_3$  be a $\frak{g}$-cocycle and $f_6$ be an $f_3$-twisted $\frak{g}$-cocycle.  In  terms of the $\frak{g}$ coboundary operator $d$ this can be stated as 
\begin{equation}
df_3 = 0 \, , \quad df_6 + f_3 \wedge f_3 = 0 \, .  
\end{equation}
We can solve this requirement with a coboundary Ansatz, $f_3 = d\rho_3   $ and  $f_6= d\rho_6 + \frac{1}{2} \rho_3 \wedge d\rho_3$, reminiscent of the way a Drinfel'd double can be constructed through an $r$-matrix.  Indeed, we find a generalised version for the Yang Baxter equation for $\rho_3$, concisely expressed as 
\begin{equation} 
\sigma_{[123],[45]}  [\rho_{123} +\rho_{234} , \rho_{145}]   =   \frac{1}{2}\left( \rho_{12345;5} +\rho_{12345;4}  \right)\, .
\end{equation}
We proposed a `quantum' relation from which this classical equation can be obtained.  This feature, and the resultant interplay between one-, three-, and six-algebras, opens up many interesting avenues for further exploration. 
 
 The construction of the EDA is closely motivated by considerations within exceptional generalised geometry.  We have shown how the EDA can be realised as a generalised Leibniz parallelisation of the exceptional generalised tangent bundle of a group manifold $G$.  The data required to construct this mean that $G$ is equipped with a 3-bracket and a 6-bracket   which invites the consideration of Nambu-Lie groups.  
 
 Now we come to solving the various constraint equations that govern the structure of the EDA.  The first thing to note is that due to the dimension, the only solutions to the fundamental identities have vanishing $f_6$ (and consequently a trivial 6-bracket on $G$).  We believe however that in higher dimension this condition is less stringent and that there will solutions for which the structure described above is exhibited in full.

 We then provided a range of examples that illustrate the various features here.    We have examples with and without Drinfel'd double subalgebras, and examples that are both of coboundary type (specified by a $\rho_3$ and $\rho_6$) and not of coboundary type.   All of the coboundary examples   presented here (and indeed in all the numerous other examples we have found) can be obtained from the procedure of $\rho$-twisting  i.e.  starting with a semi-Abelian EDA and applying an $E_{6(6)}$ transformation parametrised by the $\rho_3$ and $\rho_6$.  Despite the dimensionality induced restriction to $f_6=0$,   there are examples for which $\rho_6 \neq 0$.   We provide examples where $\rho_3$ can be parametrised in terms of a Yang-Baxter $r$-matrix for a lower dimensional algebra, as well as   where this is not the case.  
 
There are several exciting open directions here that we share in the hope that others may wish to develop them further: 
\begin{itemize}
    \item Extensions of the EDA to $E_{7(7)}$ and higher are likely to shed further light on the structures involved. As the space gets larger there is more scope to find interesting solutions. 
    \item It would be interesting to develop a more general classification of EDA solutions.
    \item One feature of the EDA is that they may admit multiple decompositions into physical spaces, and a resultant notion of duality.  Further development should go into  this very interesting aspect.
    \item Here we make some robust requirements that result in structures compatible with maximally supersymmetric gauged supergravities.  It would likely be interesting to see how the requirements of the EDA can be consistently relaxed to lower supersymmetric settings, for example using \cite{Malek:2017njj}. 
    \item On a mathematical note perhaps the most intriguing area of all is to develop the `quantum' equivalent of the classical EDA proposed here.   
\end{itemize}
 
\section{Acknowledgements}
EM is supported by ERC Advanced Grant “Exceptional Quantum Gravity” (Grant No.740209). YS is supported by JSPS Grant-in-Aids for Scientific Research (C) 18K13540 and (B) 18H01214. DCT is supported by The Royal Society through a University Research Fellowship {\em Generalised Dualities in String Theory and Holography} URF 150185 and in part by STFC grant ST/P00055X/1;   FWO-Vlaanderen through the project G006119N;  and by the Vrije Universiteit Brussel through the Strategic Research Program ``High-Energy Physics''. DCT thanks Chris Blair collaboration on related work and Tim Hollowood for informative communications.   

\appendix 
\section{  $E_{6(6)}\times \mathbb{R}^+$ Algebra and conventions}
The matrix representations of the $E_{n(n)}$ generators $(t_\alpha)_A{}^B$ in the $\Rone$-representation are
\begin{equation}
\begin{split}
 K^c{}_d &\equiv \begin{pmatrix}
 \delta_a^c\,\delta_d^b & 0 & 0 \\
 0 & - 2\,\delta_{[b_1|d|}^{a_1a_2}\,\delta_{b_2]}^{c} & 0 \\
 0 & 0 & -5\,\delta_{[b_1\ldots b_4|d|}^{a_1\ldots a_5}\,\delta_{b_5]}^{c} 
 \end{pmatrix} + \frac{\delta_d^c}{9-n}\,\mathbf{1}\,, 
\\
 R^{c_1c_2c_3} &\equiv \begin{pmatrix}
 0 & -\frac{3!\,\delta_{a b_1b_2}^{c_1c_2c_3}}{\sqrt{2!}} & 0 \\
 0 & 0 & -\frac{5!\,\delta^{a_1a_2c_1c_2c_3}_{b_1\ldots b_5}}{\sqrt{5!\,2!}} \\
 0 & 0 & 0 
 \end{pmatrix} , \quad
 R^{c_1\ldots c_6} \equiv \begin{pmatrix}
 0 & 0 & -\frac{6!\,\delta_{ab_1\ldots b_5}^{c_1\ldots c_6}}{\sqrt{5!}} \\
 0 & 0 & 0 \\
 0 & 0 & 0 
\end{pmatrix} ,
\\
 R_{c_1c_2c_3} &\equiv \begin{pmatrix}
 0 & 0 & 0 \\
 \frac{3!\,\delta^{b a_1a_2}_{c_1c_2c_3}}{\sqrt{2!}} & 0 & 0 \\
 0 & \frac{5!\,\delta_{b_1b_2 c_1c_2c_3}^{a_1\ldots a_5}}{\sqrt{2!\,5!}} & 0 
 \end{pmatrix} , \quad
 R_{c_1\ldots c_6} \equiv \begin{pmatrix}
 0 & 0 & 0 \\
 0 & 0 & 0 \\
 \frac{6!\,\delta^{ba_1\ldots a_5}_{c_1\ldots c_6}}{\sqrt{5!}} & 0 & 0 
\end{pmatrix} ,
\end{split}
\end{equation}
and that of the $\mathbb{R}^+$ generator is simply $(t_0)_A{}^B=-\delta_A^B$\,.

The non-vanishing components of symmetric invariant tensor of $E_{6(6)}$, $d^{ABC}$, are
\begin{align}
  d^{a}{}_{b_1b_2c_1\ldots c_5} = \frac{2!\,\delta^{a}_{[b_1}\,\epsilon_{b_2]c_1\ldots c_5}}{\sqrt{10}} \,,\qquad 
  d_{a_1a_2b_1b_2c_1c_2} = \frac{\epsilon_{a_1a_2b_1b_2c_1c_2}}{\sqrt{10}}\,.
\end{align}
and those  of $d_{ABC}$ are given by the same with indices in opposite positions.  Using this we have the useful tensor  given by
\begin{align} 
\eta_{AB}{}^{\mathcal{C}}\equiv \sqrt{10}\,d_{ABD}\,k^{D\mathcal{C}}\, , 
\end{align}  
in which, $k^{A\mathcal{B}}$ connects $R_1=\bf{27}$ and $R_2=\overline{\bf{27}}$,
\begin{align}
 (k^{A\mathcal{B}}) \equiv \begin{pmatrix}
 0 & 0 & \frac{\delta^a_b\,\epsilon_{b_1\cdots b_6}}{\sqrt{6!}}\\
 0 & -\frac{\epsilon_{a_1a_2 b_1\cdots b_4}}{\sqrt{2!\,4!}} & \\
 \frac{\epsilon_{a_1\cdots a_5b}}{\sqrt{5!}} & 0 & 0
\end{pmatrix},\qquad 
 (k_{\mathcal{A}B}) \equiv \begin{pmatrix}
 0 & 0 & \frac{\epsilon^{b_1\cdots b_5a}}{\sqrt{5!}} \\
 0 & -\frac{\epsilon^{b_1b_2 a_1\cdots a_4}}{\sqrt{2!\,4!}} & \\
 \frac{\delta_b^a\,\epsilon^{a_1\cdots a_6}}{\sqrt{6!}} & 0 & 0
\end{pmatrix}\, .
\end{align}
This plays a similar role to that of the $O(d,d)$ invariant inner product of generalised geometry and indeed can be used to construct the Y-tensor \cite{Berman:2012vc} given by $Y^{AB}_{CD} = \eta^{AB}{}_{\mathcal{E}}\,\eta_{CD}{}^{\mathcal{E}}= 10\, d^{ABE}\,d_{CDE}$ that is ubiquitous in exceptional field theory/generalised geometry. 

The invariance of $\eta_{AB}{}^{\cC}$ under $E_{n(n)}\times \mathbb{R}^+$ symmetry can be expressed as
\begin{align}
 (t_{\hat{\alpha}})_A{}^D\,\eta_{DB}{}^{\cC} + (t_{\hat{\alpha}})_{B}{}^D\,\eta_{AD}{}^{\cC} = \eta_{AB}{}^{\CD}\,(t_{\hat{\alpha}})_\CD{}^\cC \,. 
\label{eq:invariance}
\end{align}
From this relation, we find the matrix representation $(t_{\alpha})_{\cA}{}^{\cB}$ in the $\Rtwo$-representation is
\begin{equation}
    \begin{split}
 K^c{}_d &\equiv \begin{pmatrix}
 -\delta_d^a\,\delta_b^c & 0 & 0 \\
 0 & -4 \delta_{d e_1e_2e_3}^{a_1\ldots a_4}\,\delta_{b_1\ldots b_4 }^{c e_1 e_2 e_3} & 0 \\
 0 & 0 & - 6 \delta_{d e_1 \ldots e_5}^{a_1 \dots a_6}\,\delta_{b_1 \ldots b_6}^{c e_1 \ldots e_5 }\,\delta_{b'}^{a'} - \delta_{b_1 \ldots b_6}^{a_1 \ldots a_6 }\,\delta_d^{a'}\,\delta^c_{b'}
 \end{pmatrix} + \frac{2\,\delta_d^c}{9-n}\,\mathbf{1}\,, 
\\
 R^{c_1 c_2 c_3} &\equiv \begin{pmatrix}
 0 & -\sqrt{4!}\delta^{a c_1 c_2c_3}_{b_1 \ldots b_4} & 0 \\
 0 & 0 & - 3\sqrt{30}  \delta_{b_1\ldots b_6}^{a_1\ldots a_4 d_1 d_2  }\,\delta_{b'd_1 d_2}^{c_1 c_2 c_3} \\
 0 & 0 & 0 
 \end{pmatrix} , \quad
 R^{c_1 \ldots c_6} \equiv \begin{pmatrix}
 0 & 0 & \sqrt{6!}\delta_{b_1\ldots b_6}^{c_1 \ldots c_6}\,\delta_{b'}^{a'} \\
 0 & 0 & 0 \\
 0 & 0 & 0 
\end{pmatrix} ,
\\
 R_{c_1c_2 c_3} &\equiv \begin{pmatrix}
 0 & 0 & 0 \\
\sqrt{4!} \delta_{b c_1c_2 c_3}^{a_1 \ldots a_4} & 0 & 0 \\
 0 & 3\sqrt{30} \delta^{a_1\ldots a_6}_{b_1\ldots b_4 d_1 d_2}\,\delta^{a'd_1 d_2}_{c_1 c_2 c_3} & 0 
 \end{pmatrix} , \quad
 R_{c_1 \ldots c_6 } \equiv \begin{pmatrix}
 0 & 0 & 0 \\
 0 & 0 & 0 \\
 -\sqrt{6!} \delta^{a_1\ldots a_6 }_{c_1\ldots c_6 }\,\delta^{a'}_{b'} & 0 & 0 
\end{pmatrix} .
\end{split}
\end{equation}

In terms of these generators we can express the EDA product as,
\begin{align}
 T_A \circ T_B = \widehat{\Theta}_A{}^{\hat{\alpha}}\,(t_{\hat{\alpha}})_B{}^C\, T_B \,,
\end{align}
where $\{t_{\hat{\alpha}}\}\equiv \{t_0,\,t_\alpha\}$, $\{t_\alpha\}\equiv \{K^a{}_b,\,R^{a_1a_2a_3},\,R^{a_1\cdots a_6},\,R_{a_1a_2a_3},\,R_{a_1\cdots a_6}\}$. 
The explicit form of $\widehat{\Theta}_A{}^{\hat{\alpha}}$ is as follows:
\begin{equation}
    \begin{split}
 \widehat{\Theta}_a{}^\alpha\, t_{\alpha} &\equiv f_{ab}{}^c\, K^b{}_c + \tfrac{1}{3!}\,f_a{}^{c_1c_2c_3}\, R_{c_1c_2c_3} + \tfrac{1}{6!}\,f_a{}^{c_1\cdots c_6}\, R_{c_1\cdots c_6} \,,
\\
 \widehat{\Theta}^{a_1a_2 \alpha}\, t_{\alpha} &\equiv - f_{c_1c_2}{}^{[a_1}\, R^{a_2]c_1c_2} - f_c{}^{da_1a_2}\, K^c{}_d \,,
\\
 \widehat{\Theta}^{a_1\cdots a_5\alpha}\, t_{\alpha} &\equiv - \tfrac{5}{2}\,f_{c_1c_2}{}^{[a_1}\, R^{a_2\cdots a_5] c_1c_2} 
+ 10\,f_{c}{}^{[a_1a_2a_3}\, R^{a_4a_5] c} - f_c{}^{da_1\cdots a_5}\, K^c{}_d \,,
\\
 \widehat{\Theta}_a{}^0 &\equiv \tfrac{f_{ac}{}^c}{9-n} \,, \quad
 \widehat{\Theta}^{a_1a_2 0} \equiv - \tfrac{f_c{}^{ca_1a_2}}{9-n} \,,\quad
 \widehat{\Theta}^{a_1\cdots a_50} \equiv - \tfrac{f_c{}^{ca_1\cdots a_5}}{9-n} \,.
\end{split}
\end{equation}

We can now recast the algebra of frame fields \eqref{eq:framealg} as 
\begin{align}
 \gL_{E_A} E_{B} = - \widehat{\Theta}_A{}^{\hat{\alpha}}\,(t_{\hat{\alpha}})_B{}^C\,\,E_C \,,
\end{align}
such that making us of \eqref{eq:gL-bracket}, \eqref{eq:cE-def}, and \eqref{eq:invariance}, we can easily find that the generalized frame field in the $\Rtwo$-representation also transforms covariantly as
\begin{align}
 \gL_{E_A}\cE_{\cB} = -\widehat{\Theta}_A{}^{\hat{\alpha}}\,(t_{\hat{\alpha}})_\cB{}^\cC\, \cE_\cC \,.
\end{align}

\bibliographystyle{JHEP}
\bibliography{NewBib}

\end{document}